\documentclass[prl,reprint,amsmath,amssymb,preprintnumbers,superscriptaddress,nofootinbib]{revtex4-1}

\usepackage{epsfig,epstopdf}
\usepackage{subfigure}
\usepackage{color}

\def\beq{\begin{equation}}
\def\eeq{\end{equation}}
\def\bea{\begin{eqnarray}}
\def\eea{\end{eqnarray}}

\newcommand{\lsim}{
\mathrel{\hbox{\rlap{\hbox{\lower4pt\hbox{$\sim$}}}\hbox{$<$}}}}
\newcommand{\gsim}{
\mathrel{\hbox{\rlap{\hbox{\lower4pt\hbox{$\sim$}}}\hbox{$>$}}}}

\newcommand{\dis}[1]{\begin{equation}\begin{split}#1\end{split}\end{equation}}

\begin{document}

\preprint{CTPU-17-14}
\preprint{MAD-TH-17-03}
\title{Large gauge transformation and little group for soft photons}

\author{Yuta~Hamada}
\email{yhamada@wisc.edu}
\affiliation{Department of Physics, University of Wisconsin-Madison, Madison, WI 53706, USA}
\affiliation{KEK Theory Center, IPNS, KEK, Tsukuba, Ibaraki 305-0801, Japan}
\author{Min-Seok Seo}
\email{minseokseo57@gmail.com}
\affiliation{Center for Theoretical Physics of the Universe, Institute for Basic Science (IBS), Daejeon 34051, Korea}
\affiliation{Department of Physics, Chungnam National University,  Daejeon 34134, Korea}
\author{Gary Shiu}
\email{shiu@physics.wisc.edu}
\affiliation{Department of Physics, University of Wisconsin-Madison, Madison, WI 53706, USA}

\begin{abstract}
\noindent 
Recently, large gauge transformation (LGT), the residual gauge symmetry after gauge fixing that survives at null infinity, has drawn much attention concerning soft theorems and the memory effect.
 We point out that LGT charges in quantum electrodynamics are in fact one of non-compact generators of the two dimensional Euclidean group.
 Moreover, by comparing two equivalent descriptions of gauge transformation, we suggest that LGT 
 is simply 
 another way of describing the gauged little group for massless soft photons.
\end{abstract}
\maketitle

\subsection{Introduction}
 
 Quantum mechanically, a particle can be defined in terms of a unitary, irreducible representation of the Poincar\'e group \cite{Wigner:1939cj}.
 More concretely, by fixing the momentum of the particle in a specific frame, particle states are specified by  their discrete, finite dimensional representations of the ``little group", the subgroup of the Poincar\'e group which does not alter the momentum.
 For a massive particle, the little group is simply given by $SO(3)$, indicating particles are characterized by their spins.
 Meanwhile, the little group of a massless particle is given by the two-dimensional Euclidean group $ISO(2)$, generated by three generators $\Pi_1$, $\Pi_2$, and $J$ satisfying the closed algebra,
 \dis{[\Pi_1, \Pi_2]=0,\quad [J, \Pi_1]=i\Pi_2,\quad [J, \Pi_2]=-i\Pi_1.\label{Eq:ISO(2)}}
This algebra admits continuous and infinite dimensional representations, coming from the non-compact generators $\Pi_1$ and $\Pi_2$.
 While the non-compact generators $\Pi_{1,2}$ act on the polarization vectors as a gauge transformation \cite{Weinberg:1964ew}, their effects have been ignored simply because no continuous and infinite number of observables have been found in a fixed momentum frame.
 The common practice is to simply set $\Pi_1=\Pi_2=0$ and take the helicity $J$, the spin in the direction of momentum as an observable distinguishing different particle states.
 In that case, instead of an infinite tower of helicity states raised and lowered by $\Pi_{1,2}$, as studied in detail in Ref. \cite{Schuster:2013pxj}, we only have two helicity states.

 On the other hand, it was recently pointed out that some parts of the gauge symmetry in gravity \cite{Strominger:2013jfa} and Abelian gauge theory \cite{He:2014cra} remain as an infinite number of asymptotic symmetry even after gauge fixing (For a review, see Ref. \cite{Strominger:2017zoo}).
 These large gauge transformations (LGTs) provide a simple explanation to long distance physics of soft gravitons and photons.
 For example, the soft theorems in Ref. \cite{Weinberg:1965nx} can be understood as Ward identities of LGTs at null infinity of asymptotic flat spacetime.
 We also have infinitely degenerate vacua labelled by the number of soft photon excitations, and transition between different vacua is generated by LGT charges.
 It was also suggested that LGT can be observed through the `memory effect', the permanent change of the metric resulting from gravitational wave pulse \cite{Strominger:2014pwa}\footnote{Interestingly, the gravitational memory effect in de Sitter spacetime 
 takes a similar form as that  for flat spacetime \cite{Bieri:2015jwa, Chu:2015yua,Tolish:2016ggo, Hamada:2017gdg}
and 
 can be parametrized by a Bondi-van der Burg-Metzner-Sachs (BMS)-like supertranslation \cite{Hamada:2017gdg} despite the different asymptotics.}
  or in the case of an Abelian gauge field, the Aharonov-Bohm effect \cite{Susskind:2015hpa}.
 One intriguing implication is the possible role of LGT in the black hole information paradox, with the LGT charges being the additional `hair' distinguishing different black hole states \cite{Dvali:2015rea}.
 This suggestion has been challenged, mainly due to the separation between S-matrix elements for soft photon or graviton emission and for hard processes \cite{Gabai:2016kuf, Mirbabayi:2016axw}.
 Detailed study shows that such separation crucially depends on the conservation of LGT charge eigenvalues of degenerate vacua \cite{Gabai:2016kuf}.

 In this letter, we show that LGT charges in Abelian gauge theory for each fixed momentum direction are in fact identified with one of non-compact generators of $ISO(2)$.
  Hence, LGT charge eigenvalues are given by 
  continuous real numbers, and we can introduce another generator which commutes with the LGT generator to complete the $ISO(2)$ algebra together with the helicity operator.
 Moreover, by comparing two equivalent descriptions of the gauge transformation, 
i.e.,  in terms of a scalar function or in terms of an operator, we point out that at
null 
infinity, the transitions between degenerate vacua of different soft photon excitations generated by LGT charges might be effectively identified with the transitions between helicity $\pm1$ and $0$ states of  a single photon generated by the non-compact little group.
 We also expect that the same argument can be extended to the LGT of gravity.

\subsection{Representation of LGT on the degenerate vacua}

 We begin our discussion with a brief review of the LGT of $U(1)$ Abelian gauge theory such as quantum electrodynamics (QED).
 Under some specific choice of gauge, while all gauge degrees of freedom are eliminated globally, a part of gauge symmetries emerge at null infinity after we impose appropriate boundary conditions.
 For example, suppose we take the Lorenz gauge and the boundary conditions
 {\small
 \dis{\lim_{r\to\infty}A_u={\cal O}(r^{-1}), 
 \lim_{r\to\infty}A_r={\cal O}(r^{-2}), 
 \lim_{r\to\infty}A_{z/\bar{z}}={\cal O}(1), \nonumber}
 }where $u \equiv t-r$ is the retarded time and the angular variables are parametrized by $z=\tan(\theta/2)e^{i\phi}$ and ${\bar z}$, in terms of which 
 \footnote{With this choice of coordinate system, $(u, r, z, \bar{z})$, the flat spacetime metric is given by
 \dis{ ds^2=-du^2-2du dr+2r^2\gamma_{z{\bar z}} dz d\bar{z},\quad \gamma_{z{\bar z}}=\frac{2}{(1+z\bar{z})^2}.\nonumber}} 
 \dis{&x=(u+r, r \hat{\mathbf{x}}_{z\bar{z}}),
 \\
 & \hat{\mathbf{x}}_{z\bar{z}}=\frac{1}{1+z\bar{z}}(z+\bar{z},-i(z-\bar{z}),1-z\bar{z}).} 
 The LGT corresponds to the gauge transformation $A_\mu \to A_\mu +\partial_\mu \varepsilon$ arising from the scalar field $\varepsilon$ satisfying $\nabla^2 \varepsilon=0$.
The gauge transformation compatible with the above boundary conditions, $\varepsilon$ is given by a function of $z$ and $\bar{z}$ only, say, $\varepsilon(z,\bar{z})$ at $r\to\infty$. 
 Such emergent gauge symmetry at large scale  is not a gauge symmetry in an exact sense because it is not defined over all spacetime points and the transformation is generated by an infinite number of generators, rather than a single generator that can be continuously deformed to the identity. 
 The charge  measured at future null infinity is given by the spatial integration of the Noether current $J^\mu=\partial_\nu(F^{\mu\nu}\varepsilon)$ \cite{He:2014cra, Gabai:2016kuf},
 \dis{Q_\varepsilon=\int d^2z\, \varepsilon(z,\bar{z})\Big[&\gamma_{z\bar{z}} r^2 \lim_{u,r \to\infty}F_{ru}
 \\
 +&\int du\lim_{r\to\infty} \partial_u(\partial_z A_{\bar z}+\partial_{\bar z}A_z)\Big].\label{Eq:asycharge0}}
The first term is known as the ``hard part", that would have been the electric charge for a constant $\varepsilon$.
Since we are interested in the vacuum where the field strength vanishes in the far future, we ignore it.
On the other hand, the ``soft part", the second term corresponds to  the LGT generators, implying that charge conservation is accomplished after taking the soft photon contributions into account under the retarded time coordinate.
From the chiral behavior of the gauge field around null infinity 
{\small
\dis{A_{z/\bar{z}}=-i\sqrt{\gamma_{z\bar{z}}}\int_0^\infty\frac{d \omega}{8\pi^2}(a_\pm(\omega \hat{\mathbf{x}}_{z\bar{z}})e^{-i\omega u}-a_\mp^\dagger(\omega \hat{\mathbf{x}}_{z\bar{z}}) e^{i\omega u}),\label{Eq:Az/zb}}
}
the soft part of LGT generator is given by
 \dis{Q_\varepsilon=\int d^2z &\lim_{\omega \to 0} \frac{\omega \sqrt{\gamma_{z\bar{z}}}}{8\pi}
 \\
& \times \Big[(\partial_z \varepsilon(z, \bar{z})a_+(\omega \hat{\mathbf{x}}_{z\bar{z}})+\partial_{\bar z}\varepsilon(z,\bar{z})a^{ \dagger}_+(\omega \hat{\mathbf{x}}_{z\bar{z}}))
\\
&+(\partial_{\bar z} \varepsilon(z, \bar{z})a_-(\omega \hat{\mathbf{x}}_{z\bar{z}})+\partial_{z}\varepsilon(z,\bar{z})a^\dagger_-(\omega \hat{\mathbf{x}}_{z\bar{z}}))\Big],\label{Eq:asycharge}}
 where soft photons with polarizations $\epsilon_+^\mu=\frac{1}{\sqrt2}(\bar{z},1,-i,-\bar{z})$ and
 $\epsilon_-^\mu=\frac{1}{\sqrt2}(z,1,i,-z)$ are created and annihilated by $a^\dagger_\pm$ and $a_\pm$, respectively. 
 
  Then, the action of $Q_\varepsilon$ corresponds to transition from $n\text{--}$soft photon excitation state into $(n-1)\text{--}$ and $(n+1)\text{--}$soft photon excitation states.
  By redefinition and rescaling, we can express the charge operator in the form of
  \footnote{ In the presence of both infrared (IR) and unltraviolet (UV) cutoffs, we may regard the continuous parameter $z$ as a discrete parameter.
 In this case, the Hilbert space $\prod_{\lambda,z} \otimes {\cal H}_{\lambda, z}$ which we will discuss later is obviously separable.
 While the IR cutoff is essential to obtain a finite S-matrix elements after the IR divergence cancellation \cite{Bloch:1937pw}, the theory becomes insensitive to the UV cutoff after renormalization. 
}
  \dis{Q=\frac{1}{\sqrt{2}}\int d^2z\sum_{\lambda=+,-} (\alpha_{ \lambda,z}+\alpha_{\lambda, z}^\dagger),}
  where the creation and annihilation operators satisfy
  \dis{&[\alpha_{\lambda, z}, \alpha_{\lambda',z'}^\dagger]=\delta_{\lambda\lambda'}\delta^2(z-z'),
  \\
  &[\alpha_{\lambda, z}, \alpha_{\lambda',z'}]=[\alpha_{\lambda, z}^\dagger, \alpha_{\lambda',z'}^\dagger]=0.}

 In the presence of symmetry, the symmetry generator $Q$ commutes with the Hamiltonian $H$, and states can be described in terms of a common eigenbasis of $Q$ and $H$.   
  However, the non-excitation state in the Fock space $|0\rangle$ is not appropriate for a vacuum for LGT in this sense because it is not a LGT eigenstate.
  Rather, we need to find the vacuum as an eigenstate of LGT.
 Indeed, as soft photon excitations have (almost) zero energy-momentum, we can say that soft photon excitation states are (almost) degenerate with $|0\rangle$, and the vacuum as an eigenstate of $Q$ can be constructed as the superposition of $|0\rangle$ and (infinitely many) excitations of soft photons.

 More concretely, let us pick up one Hilbert space ${\cal H}_{\lambda, z}$ labelled by one specific mode $(\lambda, z)$.  
 The full Hilbert space is understood as a direct product $\prod_{\lambda,z} \otimes {\cal H}_{\lambda, z}$.
 For this specific mode, the charge operator $Q(\lambda, z)=\frac{1}{\sqrt2}( \alpha_{\lambda, z}+\alpha^\dagger_{\lambda,z})$ is represented in the zero-energy basis
 \dis{|0\rangle,~\alpha_{\lambda, z}^\dagger |0\rangle,~\cdots,~ |n ;\lambda, z\rangle=\frac{1}{\sqrt{n!}}(\alpha_{\lambda, z}^\dagger)^n|0\rangle,~\cdots,  }
by
{\small
 \dis{Q(\lambda, z)=\frac{1}{\sqrt2}\left[
\begin{array}{cccccc}
0 & 1 & 0 &0 & 0 &\cdots  \\
1 & 0 & \sqrt{2} &0 &0 &\cdots \\
0 & \sqrt{2} & 0 &\sqrt{3} &0 &\cdots  \\
0 & 0 & \sqrt{3} &0 &\sqrt{4}  &\cdots  \\
0 & 0 &0& \sqrt{4}& 0 &\cdots\\
& & \cdots & & & \cdots
\end{array}\right]\delta_{\lambda\lambda'}\delta^2(z-z').\label{Eq:Qmatrix}} 
}
This is the same representation as the coordinate operator for the harmonic oscillator. 
Hence, the eigenvalues $q$, the possible values of asymptotic charge, extend over
the infinite space of continuous real numbers.
Given an eigenvalue $q$ of $Q(\lambda, z)$, the vacuum $|q;\lambda, z\rangle$ is expanded in terms of $n$ soft photon excitation states as
\dis{&|q ;\lambda, z\rangle=\sum_{n}|n ; \lambda, z\rangle\langle n ; \lambda, z|q;\lambda, z \rangle,
\\
&\langle q ; \lambda, z| n; \lambda, z\rangle=\Big(\frac{1}{\pi 2^n(n!)^2}\Big)^{1/2} e^{-\frac{q^2}{2}}H_n(q),
}
with $H_n(q)$ being the Hermite polynomials.

\subsection{Equivalence of LGT and the little group}

In order to understand the presence of infinitely many, continuous eigenvalues of $Q(\lambda, z)$, we recall that the coordinate operator $Q=(1/\sqrt{2})(a+a^\dagger)$ for the harmonic oscillator has a conjugate momentum $P=-(i/\sqrt{2})(a-a^\dagger)$.
In the same way, we define an operator $P(\lambda, z)$ conjugate to the charge $Q(\lambda, z)$ by  
\dis{P=\int d^2z \sum_{\lambda=+,-}P(\lambda, z), ~ P(\lambda, z)=\frac{-i}{\sqrt2}(\alpha_{\lambda,z}-\alpha_{\lambda,z}^\dagger).}
Now, the operator
\dis{J(z)=\alpha^{\dagger}_{+,z}\alpha_{+,z}-\alpha^{\dagger}_{-,z}\alpha_{-,z}}
measures the helicity of a soft photon moving along the direction $\hat{\mathbf{x}}_{z\bar{z}}$ : $+1$ for left-handed and $-1$ for right-handed helicity.
Moreover, we define two operators having continuous eigenvalues,
\dis{\Pi_1^{L}(z)&=Q(+,z)+Q(-,z)
\\
&=\frac{1}{\sqrt2}(\alpha_{+,z}+\alpha^\dagger_{+,z})+\frac{1}{\sqrt2}(\alpha_{-,z}+\alpha^\dagger_{-,z}),
\\
\Pi_2^{L}(z)&=-P(+,z)+P(-,z)
\\
&=\frac{i}{\sqrt2}(\alpha_{+,z}-\alpha^\dagger_{+,z})-\frac{i}{\sqrt2}(\alpha_{-,z}-\alpha^\dagger_{-,z}).\label{Eq:noncomp}}
Then, $J=\int d^2 z J(z)$ and $\Pi^{L}_{1,2}=\int d^2z \Pi^{L}_{1,2}(z)$
form the closed algebra of $ISO(2)$,
\dis{[\Pi^L_1, \Pi^L_2]=0,\quad [J, \Pi^L_1]=i \Pi^L_2,\quad [J, \Pi^L_2]=-i \Pi^L_1.}
We note here that both $P(\lambda, z)$ and $Q(\lambda, z)$ do not alter the momentum of soft photon parametrized by $z$  so it is consistent with the definition of the little group. 

 In fact, the close relation between gauge transformation and the $ISO(2)$ little group was pointed out in Ref. \cite{Weinberg:1964ew}, (see also Ch. 5 of Ref.  \cite{Weinberg:1995mt}) : the massless vector field is a vector {\it up to gauge transformation}.
 The action of a unitary representation of an $ISO(2)$ little group element
$W(\theta, \alpha, \beta)={\rm exp}[-i(\alpha \Pi_1+\beta \Pi_2)]{\rm exp}[-i\theta J]$
on the polarization vectors is given by
\dis{D^\mu_{~\nu}(W(\theta, \alpha, \beta))\epsilon_\pm^\nu(\mathbf{k})=e^{\pm i\theta}\Big[\epsilon_\pm^\mu(\mathbf{k})+\frac{\alpha\pm i\beta}{\sqrt{2}\omega_k}k^\mu\Big]. \label{Eq:gauge}}
Whereas the phase factor on the right-handed side obviously indicates the rotation determined by  the helicity of the photon, the gauge transformation part shows that the action of the little group on the photon induces the gauge transformation.
Indeed, the LGT action $i[Q_\varepsilon, A_{z/\bar{z}}]=\partial_{z/ \bar{z}}\varepsilon(z, \bar{z})$ corresponding to the real part of Eq. (\ref{Eq:gauge}) comes from a scalar field 
{\small
\dis{&\Lambda_1=i\int\frac{d^3 k}{(2\pi)^3 2\omega_k}\frac{\alpha}{\sqrt2}\frac{1}{\omega_k}(e^{-i k\cdot x}-e^{i k\cdot x})
\\
& \to -\int^\infty_0 \frac{d\omega_k d^2z_k}{8\pi^2r}\frac{\alpha}{\sqrt2 \omega_k}(e^{-i\omega_k u}+e^{i\omega_k u})\delta^2(z_k-z_x), \label{Eq:gaugef}}
}
at $r\to \infty$ such that
{\small 
\dis{\lim_{r\to\infty}\partial_\mu \Lambda_1 = -
\int^\infty_0& \frac{d\omega_k d^2z_k}{8\pi^2r}\frac{\alpha}{\sqrt2\omega_k}(e^{-i\omega_k u}+e^{i\omega_k u})
\\
&\times(\partial_\mu z_x\partial_{z_x}+
\partial_\mu \bar{z}_x\partial_{\bar{z}_x})
\delta^2(z_k-z_x)+\cdots.}
}
If $ISO(2)$ is gauged, i.e., $\alpha$ is allowed to have a spacetime dependence as $\alpha=4\sqrt2 \pi^2 r \varepsilon(z, \bar{z})$, 
the soft part ($\omega_k \simeq 0$) of the gauged $ISO(2)$ transformation is just given by $\partial_z\Lambda_1=\partial_z \varepsilon(z, \bar{z})$ and $\partial_{\bar{z}}\Lambda_1=\partial_{\bar{z}} \varepsilon(z, \bar{z})$.
 On the other hand, as we build up the LGT charge $Q_\varepsilon$ through $\partial_u(A_z+A_{\bar z})$ with $A_z$ and $A_{\bar z}$ given by Eq. (\ref{Eq:Az/zb}), we can also introduce two real combinations of creation and annihilation operators of photons,
 {\small
\dis{P_{2 \pm}=\sqrt{\gamma_{z\bar{z}}}\int\frac{d \omega}{8\pi^2}[a_{\pm}(\omega \hat{\mathbf{x}}_{z\bar{z}})e^{-i\omega u}+a_\pm^{\dagger}(\omega \hat{\mathbf{x}}_{z\bar{z}}) e^{i\omega u}].}
}
While both combinations $\partial_u(P_{2+}\pm P_{2-})$ are real, we should choose $\partial_u(P_{2+}- P_{2-})$ since it commutes with $\partial_u(A_z+A_{\bar z})$ and can be simultaneously diagonalized. 
 This is nothing but $\Pi^L_2$ in Eq.  (\ref{Eq:noncomp}) and  it generates the gauge transformation corresponding to the imaginary part of Eq. (\ref{Eq:gauge}) in which a relative sign was assigned to different helicities. 

 At first glance, our identification of LGT with the little group looks strange, since the LGT charges generate transitions between different vacua composed of {\it different number} of soft photons, whereas the non-compact generators $\Pi_{1,2}$ of the little group generate transitions between different polarization states in {\it a single photon} state resulting from their spin raising/lowering properties given by $[J, \Pi_\pm]=\pm \Pi_\pm$, where $\Pi_\pm=\Pi_1\pm i\Pi_2$. 
 Such different roles of the $ISO(2)$ group generators can be traced to the two different descriptions of the gauge transformation. 
 In the case of LGT, the gauge transformation generated by $Q_\Lambda$ is $i[Q_\Lambda, A_\mu]=\partial_\mu \Lambda_1$, where $\Lambda_1$ is the {\it scalar function} given by Eq. (\ref{Eq:gaugef}).
 As a result, the LGT generator $Q_\Lambda$ is linear in the creation and annihilation operators of the photon, so when it acts on a state with $n$ photons, the state becomes either an $(n-1)\text{--}$ or an $(n+1)\text{--}$ photon state, as in Eq. (\ref{Eq:Qmatrix}). 
 In contrast, the gauge transformation induced by the non-compact little group generators $\Pi_\Lambda$, $i[\Pi_{\Lambda}, A_\mu]=\partial_\mu \hat{\Lambda}_1$ comes from the {\it operator}, such as
{\small
\dis{\partial_\mu\hat{\Lambda}_1&=\int\frac{d^3 k}{(2\pi)^3 2\omega_k}\frac{\alpha}{\sqrt2}\frac{k_\mu}{\omega_k}(a_0(\mathbf{k})e^{-i k\cdot x}+a^\dagger_{0}(\mathbf{k})e^{i k\cdot x}),}
}
where the helicity zero creation (annihilation) operators $a_0^\dagger$ ($a_0$) are used.
Unlike the case for QED, we can consider representations of the little group where $\Pi_1=\Pi_2=0$ is not imposed.
In this case, the helicity index $\lambda$  can take any (half-)integer value,
and the helicity operator is given by 
\dis{J=\int d^2 z\sum_{\lambda \in \mathbb{Z}}\lambda \alpha_{\lambda, z}^\dagger \alpha_{\lambda, z}.\label{Eq:helicity}}
We can find the operators which satisfy $ISO(2)$ algebra:
 \dis{&\Pi_1=\frac12 \int d^2 z \sum_{\lambda \in \mathbb{Z}}(\alpha_{\lambda+1,z}^\dagger\alpha_{\lambda, z}+\alpha^\dagger_{\lambda-1,z}\alpha_{\lambda, z}),
 \\
 &\Pi_2=\frac{-i}{2} \int d^2z \sum_{\lambda \in \mathbb{Z}}(\alpha_{\lambda+1,z}^\dagger\alpha_{\lambda, z}-\alpha^\dagger_{\lambda-1,z}\alpha_{\lambda, z}).}
Here, the action of $\Pi_{1,2}$ is not just annihilating (creating) a particle with a given helicity $\lambda$, but {\it also} creating (annihilating) another particle with helicity $\lambda \pm 1$.

In QED, we can interpret the gauge transformation as a transition from a state with polarization vector $\epsilon^\mu_\pm$  to a state with polarization vector $k^\mu$.
Since $(\Pi_\pm)^\mu_{~\nu} \epsilon^\nu_\pm=(\Pi_\pm)^\mu_{~\nu} k^\nu=0$ in four-dimensional representations,\footnote{We can see this by taking $\Pi_1=J_2+K_1$ and $\Pi_2=-J_1+K_2$ in terms of the rotation and boost generators for $z=0$.} the action of the little group transformation on the photon is just a gauge transformation. 
The transitions between different polarization states by the little group action are summarized as follows:
\dis{(\lambda=-1)\stackrel{\Pi_+}{\longrightarrow}&(\lambda=0)\stackrel{\Pi_+}{\longrightarrow}({\rm annihilation})
\\
({\rm annihilation})\stackrel{\Pi_-}{\longleftarrow}&(\lambda=0)\stackrel{\Pi_-}{\longleftarrow}(\lambda=+1).}
 At null infinity, this gauge transformation remains as LGT.

\subsection{Summary and Outlook}

 We have shown that the LGT charges in an Abelian gauge theory for each fixed momentum are equivalent to the non-compact part of $ISO(2)$.
 We have suggested the interpretation that LGT might be identified with the little group for massless photon, where the group action is gauged. 
 This can be viewed both from an algebraic consideration and from a comparison with the action of the $ISO(2)$ group on gauge fields.
 We have found that LGT and little group transformation correspond to different ways to describe the gauge transformation.
 
 The infinitely many continuous eigenvalues of LGT charge can be used to label the degenerate vacua comprised of different number of soft photon excitations.
 This completes the proof in Ref. \cite{Gabai:2016kuf} for the separation of hard processes from soft photon states under the conservation of LGT charge.

Our result begs the question whether the LGT for gravity given by BMS supertranslations or superrotations  has the $ISO(2)$ group structure.
 Obviously, this LGT is the symmetry separated from the Poincar\'e group, to which little group belongs, and the algebra generated by LGT in gravity \cite{Barnich:2009se} is
 known to eliminate continuous spin representations once the supertranslation eigenvalues are fixed \cite{McCarthy:1972}. 
Nonetheless, if we allow supertranslations to be unfixed, we may expect the $ISO(2)$ group structure to be generated by supertranslation generators.
Indeed,the subleading fluctuation of the $(zz)$ component of the metric $\delta g_{zz}$ contributing to the gravitational memory effect has the same structure as $A_{z/\bar{z}}$, given by Eq. (\ref{Eq:Az/zb}) \cite{Strominger:2013jfa} so an algebra similar to $ISO(2)$ might be constructed.

Interestingly, it has been argued that continuous spin representations of the Poincar\'e group 
are not present in string theory constructions \cite{Font:2013hia}. It is apparent, particularly in the light cone gauge, that the excited string states and the massless states are related by the Virasoro generators and so the absence of continuous spin representations for the former would seem to imply the same for the latter. However, this argument \cite{Font:2013hia}
does not exclude the possibility of dressing an energy eigenstate with an arbitrary number of soft photons. 
The equivalence between LGT and the little group we found here provides an interpretation of these elusive continuous spin representations in quantum field theory and string theory.

\vspace{5mm}
\begin{acknowledgments}

Acknowledgments: MS thanks the String Theory and Theoretical Cosmology research group, Department of Physics at the University of Wisconsin-Madison for hospitality during his visit.
MS was supported by IBS (Project Code IBS-R018-D1).
YH is supported by the Grant-in-Aid for Japan Society for the Promotion of Science Fellows, No. 16J06151.
GS is supported in part by the DOE grant DE-FG-02-95ER40896 and the Kellett Award of the University of Wisconsin.

\end{acknowledgments}



\begin{thebibliography}{99}

\bibitem{Wigner:1939cj} 
  E.~P.~Wigner,
  Annals Math.\  {\bf 40}, 149 (1939)
  [Nucl.\ Phys.\ Proc.\ Suppl.\  {\bf 6}, 9 (1989)].
  
\bibitem{Weinberg:1964ew} 
  S.~Weinberg,
  Phys.\ Rev.\  {\bf 135}, B1049 (1964).
  
\bibitem{Schuster:2013pxj} 
  P.~Schuster and N.~Toro,
  JHEP {\bf 1309}, 104 (2013)
  [arXiv:1302.1198 [hep-th]];
  P.~Schuster and N.~Toro,
  JHEP {\bf 1309}, 105 (2013)
  [arXiv:1302.1577 [hep-th]].
  P.~Schuster and N.~Toro,
  Phys.\ Rev.\ D {\bf 91}, 025023 (2015)
  [arXiv:1404.0675 [hep-th]].
  
\bibitem{Strominger:2013jfa} 
  A.~Strominger,
  JHEP {\bf 1407}, 152 (2014)
  [arXiv:1312.2229 [hep-th]];
  T.~He, V.~Lysov, P.~Mitra and A.~Strominger,
  JHEP {\bf 1505}, 151 (2015)
  [arXiv:1401.7026 [hep-th]].
  
\bibitem{He:2014cra} 
  T.~He, P.~Mitra, A.~P.~Porfyriadis and A.~Strominger,
  JHEP {\bf 1410}, 112 (2014)
  [arXiv:1407.3789 [hep-th]];
  D.~Kapec, M.~Pate and A.~Strominger,
  arXiv:1506.02906 [hep-th].
  
\bibitem{Strominger:2017zoo} 
  A.~Strominger,
  arXiv:1703.05448 [hep-th].
  
\bibitem{Weinberg:1965nx} 
  S.~Weinberg,
  Phys.\ Rev.\  {\bf 140}, B516 (1965).
  
\bibitem{Strominger:2014pwa} 
  A.~Strominger and A.~Zhiboedov,
  JHEP {\bf 1601}, 086 (2016)
  [arXiv:1411.5745 [hep-th]].
  
\bibitem{Bieri:2015jwa} 
  L.~Bieri, D.~Garfinkle and S.~T.~Yau,
  Phys.\ Rev.\ D {\bf 94}, no. 6, 064040 (2016)
  [arXiv:1509.01296 [gr-qc]].
  
\bibitem{Chu:2015yua} 
  Y.~Z.~Chu,
  Phys.\ Rev.\ D {\bf 92}, no. 12, 124038 (2015)
  [arXiv:1504.06337 [gr-qc]].
  Y.~Z.~Chu,
  Class.\ Quant.\ Grav.\  {\bf 34}, no. 3, 035009 (2017)
  [arXiv:1603.00151 [gr-qc]].
  
\bibitem{Tolish:2016ggo} 
  A.~Tolish and R.~M.~Wald,
  Phys.\ Rev.\ D {\bf 94}, no. 4, 044009 (2016)
  [arXiv:1606.04894 [gr-qc]].
  
\bibitem{Hamada:2017gdg} 
  Y.~Hamada, M.~S.~Seo and G.~Shiu,
  arXiv:1702.06928 [hep-th].
  
\bibitem{Susskind:2015hpa} 
  L.~Susskind,
  arXiv:1507.02584 [hep-th].
  
\bibitem{Dvali:2015rea} 
  G.~Dvali, C.~Gomez and D.~L\"ust,
  Phys.\ Lett.\ B {\bf 753}, 173 (2016)
  [arXiv:1509.02114 [hep-th]];
  S.~W.~Hawking, M.~J.~Perry and A.~Strominger,
  Phys.\ Rev.\ Lett.\  {\bf 116}, no. 23, 231301 (2016)
  [arXiv:1601.00921 [hep-th]].
  
  
\bibitem{Mirbabayi:2016axw} 
  M.~Mirbabayi and M.~Porrati,
  Phys.\ Rev.\ Lett.\  {\bf 117}, no. 21, 211301 (2016)
  [arXiv:1607.03120 [hep-th]].
   
\bibitem{Gabai:2016kuf} 
  B.~Gabai and A.~Sever,
  JHEP {\bf 1612}, 095 (2016)
  [arXiv:1607.08599 [hep-th]].
 
\bibitem{Bloch:1937pw} 
  F.~Bloch and A.~Nordsieck,
  Phys.\ Rev.\  {\bf 52}, 54 (1937).

\bibitem{Weinberg:1995mt} 
  S.~Weinberg,
  ``The Quantum theory of fields. Vol. 1: Foundations,''
  Cambridge University Press, 1995.
  
\bibitem{Barnich:2009se} 
  G.~Barnich and C.~Troessaert,
  Phys.\ Rev.\ Lett.\  {\bf 105}, 111103 (2010)
  [arXiv:0909.2617 [gr-qc]];
  G.~Barnich and C.~Troessaert,
  JHEP {\bf 1112}, 105 (2011)
  [arXiv:1106.0213 [hep-th]].
  
\bibitem{McCarthy:1972} 
P. J. McCarthy, Proc. Roy. Soc. Lond., {\bf A330},517 (1972); Proc. Roy. Soc. Lond., {\bf A333},317 (1973); Proc. Roy. Soc. Lond., {\bf A335}, 301 (1973);  Proc. Roy. Soc. Lond., {\bf A351},55 (1976).
  
\bibitem{Font:2013hia} 
  A.~Font, F.~Quevedo and S.~Theisen,
  Fortsch.\ Phys.\  {\bf 62}, 975 (2014)
  [arXiv:1302.4771 [hep-th]].
  

  

 
  


\end{thebibliography}
\end{document}